\documentclass[conference]{IEEEtran}
\usepackage{mathrsfs}
\usepackage{amsmath}
\usepackage{amssymb}
\usepackage{enumerate}
\usepackage{color,xcolor}
\usepackage{graphicx}
\usepackage{subfigure}
\usepackage{pifont}

\ifCLASSINFOpdf

\else

\fi

\newtheorem{theorem}{Theorem}
\newtheorem{proposition}{Proposition}
\newtheorem{corollary}{Corollary}
\newtheorem{lemma}{Lemma}

\usepackage[square,numbers,sort&compress]{natbib}

\usepackage{ifpdf}

\begin{document}

\title{On Parameterized Gallager's First Bounds for Binary Linear Codes over AWGN Channels}

\author{\authorblockN{Xiao Ma\authorrefmark{1}, Jia Liu\authorrefmark{1}\authorrefmark{2}, and Baoming Bai\authorrefmark{3}
\authorblockA{\authorrefmark{1}Department of Electronics and Communication
Engineering, Sun Yat-sen University, Guangzhou 510006, GD, China}
\authorblockA{\authorrefmark{2}College of Comp. Sci. and Eng., Zhongkai University of Agriculture and Engineering, Guangzhou 510225, GD, China}
\authorblockA{\authorrefmark{3}State Lab. of ISN, Xidian University, Xi'an 710071, Shaanxi, China}}
Email: maxiao@mail.sysu.edu.cn, ljia2@mail2.sysu.edu.cn and
bmbai@mail.xidian.edu.cn}

\maketitle

\begin{abstract}
In this paper, nested Gallager regions with a single parameter is introduced to exploit
Gallager's first bounding technique~(GFBT). We present a necessary and sufficient condition on the optimal parameter.
We also present a sufficient condition~(with a simple
geometrical explanation) under which the optimal parameter does not
depend on the signal-to-noise ratio~(SNR).
With this general framework, three existing upper bounds are revisited, including the tangential
bound~(TB) of Berlekamp, the sphere bound~(SB) of Herzberg and Poltyrev, and the tangential-sphere bound~(TSB) of Poltyrev.
This paper also reveals that the SB of Herzberg and Poltyrev is equivalent to the SB of
Kasami~{\em et~al.}, which was rarely cited in literature.

\end{abstract}

\IEEEpeerreviewmaketitle

\section{Introduction}\label{introduction}
In most scenarios, there do not exist easy ways to compute the exact
decoding error probabilities for specific codes and ensembles.
Therefore, deriving tight analytical bounds is an important research
subject in the field of coding theory and practice. Many previously
reported upper
bounds~\cite{Berlekamp80,Kasami92,Kasami93,Sphere94,TSB94,Divsalar99,Divsalar03,Yousefi04,Yousefi04a,Mehrabian06,Ma10,Ma11},
as mentioned in~\cite{Sason06}, are based on Gallager's first
bounding technique~(GFBT)
\begin{equation}\label{GFBT}
  {\rm Pr} \{E\} \leq {\rm Pr} \{E,\underline{y}\in\mathcal{R}\} + {\rm Pr}\{\underline{y}\notin \mathcal{R}\},
\end{equation}
where $E$ denotes the error event, $\underline{y}$ denotes the
received signal vector, and $\mathcal{R}$ denotes an arbitrary
region around the transmitted signal vector. The first term in the right hand side~(RHS) of~(\ref{GFBT}) is usually bounded by the union bound, while the second term in
the RHS of ~(\ref{GFBT}) represents the probability of the event
that the received vector $\underline{y}$ falls outside the region
$\mathcal{R}$, which is considered to be decoded incorrectly even if it may not fall outside the Voronoi region~\cite{Agrell96}~\cite{Agrell98} of the transmitted codeword.

For convenience, we call~(\ref{GFBT}) {\em $\mathcal{R}$-bound}. Intuitively, the more similar the region
$\mathcal{R}$ is to the Voronoi region of the transmitted signal
vector, the tighter the $\mathcal{R}$-bound is. Therefore, both the shape and the size of the region $\mathcal{R}$ are critical to GFBT. Given the region's shape, one can optimize its size to obtain the tightest $\mathcal{R}$-bound.

Different from most existing works, where the size of $\mathcal{R}$ is optimized by setting to be zero the partial derivative of the bound with respect to a parameter~(specifying the size), we will propose in this paper an alternative method by introducing nested Gallager's regions. The main contributions of this paper are summarized as follows.

\begin{enumerate}
\item We present a necessary and sufficient condition on the optimal parameter.
\item We propose a sufficient condition~(with a simple geometrical
explanation) under which the optimal parameter does not depend on
the signal-to-noise ratio~(SNR).
\item Within the general framework based on the introduced nested Gallager's regions, we re-visit three existing upper bounds,
the tangential bound~(TB) of Berlekamp~\cite{Berlekamp80}, the sphere bound~(SB) of Herzberg and Poltyrev~\cite{Sphere94} and the
tangential-sphere bound~(TSB) of Poltyrev~\cite{TSB94}. The new derivation also reveals that the SB of Herzberg and Poltyrev
is equivalent to the SB of Kasami {\em
et~al.}~\cite{Kasami92}~\cite{Kasami93}, which was rarely cited in
literature.
\end{enumerate}

\section{The Parameterized Gallager's First Bounds}\label{sec2}

\subsection{The System Model}
Let $\mathcal{C}[n,k, d_{\min}]$ be a binary linear block code of dimension
$k$, length $n$, and minimum Hamming distance $d_{\min}$.
Suppose that a codeword $\underline{c}=(c_0,
c_1, \cdots, c_{n-1}) \in \mathcal{C}$ is modulated by binary phase
shift keying~(BPSK), resulting in a bipolar signal vector
$\underline s$ with $s_t = 1 - 2c_t$ for $0\leq t \leq n-1$.
The signal vector $\underline s$ is transmitted over an AWGN channel.
Let $\underline{y} = {\underline s} + {\underline z}$ be the
received vector, where $\underline z$ is a sample from a white
Gaussian noise process with zero mean and double-sided power
spectral density $\sigma^2$. For AWGN channels, the
maximum-likelihood decoding is equivalent to finding the nearest
signal vector $\hat{\underline s}$ to $\underline y$. Without loss
of generality, we assume that the bipolar image $\underline s^{(0)}$ of the all-zero codeword $\underline
c^{(0)}$ is transmitted.

\subsection{GFBT with Parameters}
In this subsection, we present parameterized GFBT by introducing
nested Gallager regions with parameters. To this end, let
$\{\mathcal{R}(r), r \in \mathcal{I} \subseteq\mathbb{R}\}$ be a
family of Gallager's regions with the same shape and parameterized
by $r\in \mathcal{I}$. For example, the nested regions can be chosen
as a family of $n$-dimensional spheres of radius $r \geq 0$ centered
at the transmitted codeword ${\underline s}^{(0)}$. We make the
following assumptions.

{\bf Assumptions.}
\begin{enumerate}
  \item[A1.] The regions $\{\mathcal{R}(r), r \in \mathcal{I} \subseteq\mathbb{R}\}$  are {\em nested} and their boundaries partition the whole space $\mathbb{R}^n$. That is,

       \begin{equation}\label{Assump1-1}
            \mathcal{R}(r_1) \subset \mathcal{R}(r_2)~{\rm if}~r_1  < r_2,
       \end{equation}

       \begin{equation}\label{Assump1-2}
            \partial\mathcal{R}(r_1) \bigcap \partial\mathcal{R}(r_2)=\emptyset~{\rm if}~r_1 \neq r_2,
       \end{equation}

       and

       \begin{equation}\label{Assump1-3}
            \mathbb{R}^n = \bigcup\limits_{r \in \mathcal{I}} \partial\mathcal{R}(r),
        \end{equation}
        where $\partial\mathcal{R}(r)$ denotes the boundary
        surface of the region $\mathcal{R}(r)$.

  \item[A2.] Define a functional $R: {\underline y} \mapsto r$ whenever ${\underline y} \in \partial\mathcal{R}(r)$. The randomness of the received vector $\underline y$ then induces a random variable $R$. We assume that $R$ has a probability density function~(pdf) $g(r)$.

  \item[A3.] We also assume that ${\rm Pr}\{E|{\underline y} \in \partial\mathcal{R}(r)\}$ can be upper-bounded by a computable upper bound $f_u(r)$.
\end{enumerate}

For ease of notation, we may enlarge the index set $\mathcal{I}$ to $\mathbb{R}$ by setting $g(r) \equiv 0$ for $r\notin \mathcal{I}$.
Under the above assumptions, we have the following parameterized GFBT~\footnote{Strictly speaking, we need one more assumption that $f_u(r)$ is measurable with respect to $g(r)$.}.

\begin{proposition}\label{Proposition_PGFBT}
For any $r^*\in \mathbb{R}$,
\begin{equation}\label{PGFBT}
  {\rm Pr} \{E\} \leq \int_{-\infty}^{r^*} f_u(r) g(r)~{\rm d}r + \int_{r^*}^{+\infty}g(r)~{\rm d}r.
\end{equation}
\end{proposition}
\begin{IEEEproof}
\begin{eqnarray}
  {\rm Pr} \{E\} &=&  {\rm Pr}\{E, \underline y \in \mathcal{R}(r^*)\} + {\rm Pr}\{E, \underline y \notin \mathcal{R}(r^*)\}\nonumber\\
              &\leq&  {\rm Pr}\{E, \underline y \in \mathcal{R}(r^*)\} + {\rm Pr}\{\underline y \notin \mathcal{R}(r^*)\}\nonumber\\
                 &=&  \int_{-\infty}^{r^*} f_u(r) g(r)~{\rm d}r + \int_{r^*}^{+\infty}g(r)~{\rm d}r. \nonumber
\end{eqnarray}
\end{IEEEproof}

An immediate question is how to choose $r^*$ to make the above bound as tight
as possible? A natural method is to set the derivative of~(\ref{PGFBT}) with respect to $r^*$ to be zero and then solve the
equation. In this paper, we propose an alternative method for gaining insight into the optimal parameter.

Before presenting a necessary and sufficient condition on the optimal parameter, we need emphasize that
the computable bound $f_u(r)$ may exceed one. We also assume that $f_u(r)$ is non-trivial, i.e., there exists some $r$ such that $f_u(r) \leq 1$. For example, $f_u(r)$ can be taken as the union bound conditional on ${\underline y} \in \partial\mathcal{R}(r)$.

\begin{theorem}~\label{Theorem_r1}
Assume that $f_u(r)$ is a non-decreasing and continuous function of $r$. Let $r_1$ be a parameter that minimizes the upper bound as shown in~(\ref{PGFBT}). Then $r_1 = \sup\{r\in\mathcal{I}\}$ if $f_u(r) < 1$ for all $r \in \mathcal{I}$; otherwise, $r_1$ can be taken as any solution of $f_u(r) = 1$.
Furthermore, if $f_u(r)$ is strictly increasing in an interval $[r_{\min}, r_{\max}]$ such that $f_u(r_{\min}) < 1$ and $f_u(r_{\max}) > 1$, there exists a unique $r_1 \in [r_{\min}, r_{\max}]$ such that $f_u(r_1) = 1$.
\end{theorem}

\begin{IEEEproof}
The second part is obvious since the function $f_u(r)$ is strictly increasing and continuous, which is helpful for solving numerically the equation $f_u(r) = 1$.

To prove the first part, it suffices to prove that neither $r_0 < \sup\{r\in\mathcal{I}\}$ with $f_u(r_0) < 1$ nor $r_2$ with $f_u(r_2) > 1$ can be optimal.

Let $r_0 < \sup\{r\in\mathcal{I}\}$ such that $f_u(r_0) < 1$. Since $f_u(r)$ is continuous and $r_0 < \sup\{r\in\mathcal{I}\}$, we can find $\mathcal{I}\ni r' > r_0$ such that $f_u(r') < 1$. Then we have

\begin{eqnarray}
\lefteqn{\int_{-\infty}^{r_0} f_u(r) g(r)~{\rm d}r + \int_{r_0}^{+\infty}g(r)~{\rm d}r} \nonumber\\
&\!\!\!\!\!\!\!\!\!\!\! \!\!\!=&\!\!\!\!\! \!\!\!\int_{-\infty}^{r_0} f_{u}(r)g(r)~{\rm d}r +\!\!\!
\int_{r_0}^{r'} g(r)~{\rm d}r + \int_{r'}^{+\infty}g(r)~{\rm d}r \nonumber \\
&\!\!\!\!\!\!\!\!\!\!\! \!\!\!>&\!\!\!\!\! \!\!\!
\int_{-\infty}^{r_0} \!\!f_{u}(r) g(r)~{\rm d}r + \int_{r_0}^{r'}\!f_{u}(r)g(r)~{\rm d}r + \int_{r'}^{+\infty}\!\!\!\!\!g(r)~{\rm d}r \nonumber \\
&\!\!\!\!\!\!\!\!\!\!\! \!\!\!=&\!\!\!\!\!
\!\!\!\int_{-\infty}^{r'}f_{u}(r)g(r) ~{\rm d}r +
\int_{r'}^{+\infty}g(r)~{\rm d}r, \nonumber
\end{eqnarray}
where we have used the fact that $f_u(r) < 1$ for $r\in [r_0, r']$. This shows that $r'$ is better than $r_0$.

Suppose that $r_2$ is a parameter such that $f_u(r_2) > 1$. Since $f_u(r)$ is continuous and non-trivial, we can find $r_1 < r_2$ such that $f_u(r_1) = 1$. Then we have
\begin{eqnarray}
\lefteqn{\int_{-\infty}^{r_2} f_u(r) g(r)~{\rm d}r + \int_{r_2}^{+\infty}g(r)~{\rm d}r} \nonumber\\
&\!\!\!\!\!\!\!\!\!\!\! \!\!\!=&\!\!\!\!\! \!\!\!\int_{-\infty}^{r_{1}} f_{u}(r)g(r)~{\rm d}r +\!\!\!
\int_{r_{1}}^{r_2} f_{u}(r)g(r)~{\rm d}r + \int_{r_2}^{+\infty}g(r)~{\rm d}r \nonumber \\
&\!\!\!\!\!\!\!\!\!\!\! \!\!\!>&\!\!\!\!\! \!\!\!
\int_{-\infty}^{r_1} f_{u}(r) g(r)~{\rm d}r + \int_{r_1}^{r_2}g(r)~{\rm d}r + \int_{r_2}^{+\infty}\!\!\!\!\!g(r)~{\rm d}r \nonumber \\
&\!\!\!\!\!\!\!\!\!\!\! \!\!\!=&\!\!\!\!\!
\!\!\!\int_{-\infty}^{r_1}f_{u}(r)g(r) ~{\rm d}r +
\int_{r_1}^{+\infty}g(r)~{\rm d}r, \nonumber
\end{eqnarray}
where we have used a condition that $f_u(r) > 1$ for $r\in (r_1, r_2]$, which can be fulfilled by choosing $r_1$ to be the maximum solution of $f_u(r) = 1$. This shows that $r_1$ is better than $r_2$.
\end{IEEEproof}

\begin{corollary}\label{corollary_r_1_SNR}
Let $f_u(r)$ be a non-decreasing and continuous function of $r$. If $f_u(r)$ does not depend on the SNR, then the optimal parameter $r_1$ minimizing the upper bound~(\ref{PGFBT}) does not depend on the SNR, either.
\end{corollary}
\begin{IEEEproof}
It is an immediate result from Theorem~\ref{Theorem_r1}.
\end{IEEEproof}

Theorem~\ref{Theorem_r1} requires $f_u(r)$ to be a non-decreasing and continuous function of $r$, which can be fulfilled for several well-known bounds. Without such a condition, we may use the following more general theorem.

\begin{theorem}~\label{Theorem_general}
For any measurable subset $\mathcal{A} \subset \mathcal{I}$, we have
\begin{equation}\label{GeneralGFBT0}
    {\rm Pr}\{E\} \leq \int_{r\in \mathcal{A}} f_u(r)g(r)~{\rm d}r + \int_{r\notin \mathcal{A}} g(r)~{\rm d}r.
\end{equation}
Within this type, the tightest bound is
\begin{equation}\label{GeneralGFBT1}
    {\rm Pr}\{E\} \leq \int_{r\in \mathcal{I}_0} f_u(r)g(r)~{\rm d}r + \int_{r\notin \mathcal{I}_0} g(r)~{\rm d}r,
\end{equation}
where $\mathcal{I}_0 = \{r\in \mathcal{I}| f_u(r) < 1\}$. Equivalently, we have
\begin{equation}\label{GeneralGFBT2}
    {\rm Pr}\{E\} \leq \int_{r\in \mathcal{I}} \min\{f_u(r), 1\}g(r)~{\rm d}r.
\end{equation}
\end{theorem}

\begin{IEEEproof} Let $\mathcal{G} = \bigcup_{r\in \mathcal{A}}\partial \mathcal{R}(r)$, we have
\begin{eqnarray*}
    {\rm Pr}\{E\} &\leq&  {\rm Pr}\{E, {\underline y} \in \mathcal{G}\} + {\rm Pr}\{{\underline y} \notin \mathcal{G}\}\\
     &=& \int_{r\in \mathcal{A}} f_u(r)g(r)~{\rm d}r + \int_{r\notin \mathcal{A}} g(r)~{\rm d}r.
\end{eqnarray*}
Define $\mathcal{A}_0 = \{r\in \mathcal{A}| f_u(r) < 1\}$ and $\mathcal{A}_1 = \{r\in \mathcal{A}| f_u(r) \geq 1\}$. Similarly, define
$\mathcal{B}_0 = \{r\notin \mathcal{A}| f_u(r) < 1\}$ and $\mathcal{B}_1 = \{r\notin \mathcal{A}| f_u(r) \geq 1\}$.
Noticing that
\begin{eqnarray*}
\int_{r\in \mathcal{A}} f_u(r)g(r)~{\rm d}r &\geq& \int_{r\in \mathcal{A}_0} f_u(r)g(r)~{\rm d}r  + \int_{r\in \mathcal{A}_1} g(r)~{\rm d}r\\
\int_{r\notin \mathcal{A}} g(r)~{\rm d}r &\geq& \int_{r\in \mathcal{B}_0} f_u(r)g(r)~{\rm d}r  + \int_{r\in \mathcal{B}_1} g(r)~{\rm d}r,
\end{eqnarray*}
we have
\begin{eqnarray*}
& &\int_{r\in \mathcal{A}} f_u(r)g(r)~{\rm d}r + \int_{r\notin \mathcal{A}} g(r)~{\rm d}r\\
&\geq& \int_{r\in \mathcal{A}_0\bigcup \mathcal{B}_0} f_u(r)g(r)~{\rm d}r + \int_{r\in \mathcal{A}_1\bigcup \mathcal{B}_1} g(r)~{\rm d}r\\
&=&\int_{r\in \mathcal{I}_0} f_u(r)g(r)~{\rm d}r + \int_{r\notin \mathcal{I}_0} g(r)~{\rm d}r\\
&=& \int_{r\in \mathcal{I}} \min\{f_u(r), 1\}g(r)~{\rm d}r.
\end{eqnarray*}

\end{IEEEproof}

\subsection{Conditional Pair-Wise Error Probabilities}
Let $\underline c^{(1)}$ denote a codeword
of Hamming weight $d \geq 1$ with bipolar image $\underline
s^{(1)}$.
The pair-wise error probability conditional on the event $\{{\underline y} \in \partial \mathcal{R}(r)\}$, denoted by $p_2(r, d)$, is
\begin{eqnarray}\label{Cpairwise}
    p_2(r, d) &=& {\rm Pr}\left\{\|\underline{y}-\underline s^{(1)}\|\leq \|\underline{y}-\underline
  s^{(0)}\| \mid \underline{y}\in \partial\mathcal{R}(r)\right\}\nonumber\\
  &=& \frac{\int_{\|\underline{y}-\underline s^{(1)}\| \leq \|\underline{y}-\underline
  s^{(0)}\|,~~\underline{y}\in \partial\mathcal{R}(r)}f(\underline y)~{\rm d}{\underline y}}{\int_{\underline{y}\in \partial\mathcal{R}(r)}f(\underline y)~{\rm d}{\underline y}},
\end{eqnarray}
where $f({\underline y})$ is the pdf of $\underline y$. Noticing that, different from the unconditional pair-wise error probabilities, $p_2(r, d)$ may be zero for some $r$.

We have the following lemma.
\begin{lemma}\label{LemmaCpairwise}
Suppose that, conditional on $\underline{y}\in \partial\mathcal{R}(r)$, the received vector $\underline y$ is uniformly distributed over
$\partial\mathcal{R}(r)$. Then the conditional pair-wise error probability $p_2(r, d)$ does not depend on the SNR.
\end{lemma}
\begin{IEEEproof} Since $f(\underline y)$ is constant for ${\underline y} \in \partial\mathcal{R}(r)$, we have, by canceling $f(\underline y)$ from both the numerator and the denominator of $(\ref{Cpairwise})$,
\begin{equation}\label{Cpairwise1}
    p_2(r, d) = \frac{\int_{\|\underline{y}-\underline s^{(1)}\| \leq \|\underline{y}-\underline
  s^{(0)}\|, \underline{y}\in \partial\mathcal{R}(r)}~{\rm d}{\underline y}}{\int_{\underline{y}\in \partial\mathcal{R}(r)}~{\rm d}{\underline y}},
\end{equation}
which shows that the conditional pair-wise error probability can be represented as a ratio of two ``surface area" and hence does not depend on the SNR.
\end{IEEEproof}


\begin{theorem}\label{Theorem_r_1_SNR}
Let $f_u(r)$ the conditional union bound
\begin{equation}\label{CUnionbound}
    f_u(r) = \sum_{1\leq d \leq n} A_d p_2(r, d),
\end{equation}
where $\{A_d, 1\leq d \leq n\}$ is the weight distribution of the code $\mathcal{C}$.
Suppose that, conditional on $\underline{y}\in \partial\mathcal{R}(r)$, the received vector $\underline{y}$ is uniformly distributed over $\partial\mathcal{R}(r)$. If $f_u(r)$ is a non-decreasing and continuous function of $r$, then the optimal parameter $r_1$ minimizing the bound~(\ref{PGFBT}) does not depend on SNR but only on the weight spectrum of the code.
\end{theorem}
\begin{IEEEproof}
From Lemma~\ref{LemmaCpairwise}, we know that $f_u(r)$ does not depend on the SNR. From Corollary~\ref{corollary_r_1_SNR}, we know that $r_1$ does not depend on the SNR.

More generally, without the condition that $f_u(r)$ is a non-decreasing and continuous function of $r$, the optimal interval $\mathcal{I}_0$ defined in Theorem~\ref{Theorem_general} does not depend on the SNR, either.

\end{IEEEproof}

\section{Single-Parameterized Upper Bounds Revisited}\label{sec3}

Without loss of generality, we assume that the code $\mathcal{C}$ has at least three non-zero codewords, i.e., its dimension $k>1$.
Let ${\underline c}^{(1)}$~(with bipolar image ${\underline s}^{(1)}$) be a codeword of Hamming weight $d$. The Euclidean distance between $\underline s^{(0)}$ and $\underline s^{(1)}$ is $\delta_d = 2\sqrt{d}$.

\subsection{The Sphere Bound Revisited}
\begin{figure}
\centering
  \includegraphics[width=6cm]{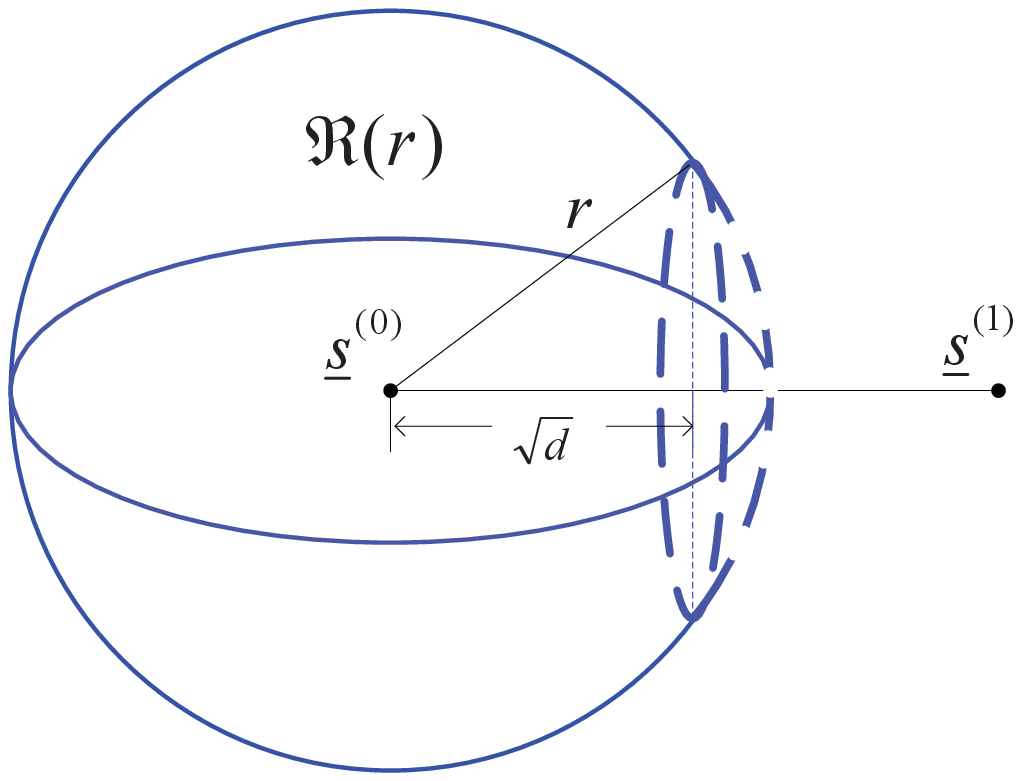}\\
  \caption{The geometric interpretation of the SB.}\label{Fig_SB}\end{figure}

\subsubsection{Nested Regions}

The SB chooses the nested regions to be a family of $n$-dimensional spheres centered at the transmitted signal vector, that is, $\mathcal{R}(r) = \{\underline{y}\mid \|\underline{y}-\underline s^{(0)}\| \leq r\}$, where $r\geq 0$ is the parameter.

\subsubsection{Probability Density Function of the Parameter}

The pdf of the parameter is
\begin{equation}\label{SB_g}
g(r) = \frac{2r^{n-1}e^{-\frac{r^{2}}{2\sigma^{2}}}}{2^{\frac{n}{2}}\sigma^{n}\Gamma(\frac{n}{2})},~~r\geq 0.
\end{equation}

\subsubsection{Conditional Upper Bound}
The SB chooses $f_u(r)$ to be the conditional union bound. Given
that $||{\underline y} - {\underline s}^{(0)}|| = r$, $\underline y$
is uniformly distributed over $\partial\mathcal{R}(r)$. Hence the
conditional pair-wise error probability $p_2(r, d)$ does not depend
on the SNR and can be evaluated as the ratio of the surface area of
a spherical cap to that of the whole sphere, as shown in
Fig.~\ref{Fig_SB}. That is,
\begin{equation}\label{SB_f2}
    \!\!\!\!\!p_2(r, d)\!\! = \!\!\left\{\begin{array}{rl}
                                         \frac{\Gamma(\frac{n}{2})}{\sqrt{\pi}~\Gamma(\frac{n-1}{2})}\int_{0}^{\arccos(\frac{\sqrt{d}}{r})}\sin^{n-2}
\phi~{\rm d}\phi, & r > \sqrt{d} \\
                                           0, & r \leq \sqrt{d}
                                         \end{array}\right.,
\end{equation}
which is a non-decreasing and continuous function of $r$ such that $p_2(0, d) = 0$ and $p_2(+\infty, d) = 1/2$. Therefore, the conditional union bound
\begin{equation}\label{SB_f_u}
    f_u(r)=\sum_{1\leq d \leq n} A_d p_2(r, d)
\end{equation}
is also an non-decreasing and continuous function of $r$ such that $f_u(0) = 0$ and $f_u(+\infty) \geq 3/2$. Furthermore, $f_u(r)$ is a strictly increasing function in the interval $[\sqrt{d_{\min}}, +\infty)$ with $f_u(\sqrt{d_{\min}}) = 0$. Hence there exists a unique $r_1$ satisfying

\begin{equation}\label{SB_opt}
    \sum_{1\leq d \leq n} A_d p_2(r, d) = 1,
\end{equation}
which is equivalent to that given in~\cite[(3.48)]{Sason06} by noticing that $p_2(r, d) = 0$ for $d > r^2$.

\subsubsection{Equivalence}

The SB can be written as
\begin{eqnarray}\label{SB}
  {\rm Pr} \{E\}&\leq& \int_{0}^{r_{1}}f_{u}(r) g(r)  ~{\rm d}r +
\int_{r_{1}}^{+\infty}g(r)~{\rm d}r\nonumber\\
&=& \int_{0}^{+\infty} \min\{f_{u}(r), 1\} g(r) ~{\rm d}r,
\end{eqnarray}
where $g(r)$ and $f_u(r)$ are given in~(\ref{SB_g})
and~(\ref{SB_f_u}), respectively. The optimal parameter $r_{1}$ is
given by solving the equation~(\ref{SB_opt}), which does not depend
on the SNR. It can be seen that~(\ref{SB}) is exactly the sphere
bound of Kasami {\em et al}~\cite{Kasami92}\cite{Kasami93}. It can
also be proved that~(\ref{SB}) is equivalent to that given
in~\cite[(3.45)-(3.48)]{Sason06}. Firstly, we have shown that the optimal radius $r_1$
satisfies~(\ref{SB_opt}), which is equivalent to that given
in~\cite[(3.48)]{Sason06}. Secondly, by changing variables, $z_{1} =
r\cos\phi$ and $y = r^{2}$, it can be verified that~(\ref{SB}) is
equivalent to that given in~\cite[Sec.3.2.5]{Sason06}.

\subsection{The Tangential Bound Revisited}

The AWGN sample $\underline z$ can be separated by projection as a radial component $z_{\xi_1}$ and $n-1$ tangential~(orthogonal) components $\{z_{\xi_i}, 2\leq i \leq n\}$. Specifically, we set $z_{\xi_1}$ to be the inner product of $\underline z$ and $-{\underline s}^{(0)}/\sqrt{n}$. When considering the pair-wise error probability, we assume that $z_{\xi_2}$ is the component that lies in the plane determined by ${\underline s}^{(0)}$ and ${\underline s}^{(1)}$.

\subsubsection{Nested Regions}

The TB chooses the nested regions to be a family of half-spaces $Z_{\xi_1} \leq z_{\xi_1}$, where $z_{\xi_1} \in \mathbb{R}$ is the parameter.

\subsubsection{Probability Density Function of the Parameter}

The pdf of the parameter is
\begin{equation}\label{TB_g}
    g(z_{\xi_1}) = \frac{1}{\sqrt{2\pi}\sigma}e^{-\frac{z_{\xi_1}^2}{2\sigma^2}}.
\end{equation}

\subsubsection{Conditional Upper Bound}
The TB chooses $f_u(z_{\xi_1})$ to be the conditional union bound. Given that $Z_{\xi_1} = z_{\xi_1}$, the conditional pair-wise error probability is given by
\begin{equation}\label{TB_f2}
    p_2(z_{\xi_1}, d) =\int_{\frac{\sqrt{d}(\sqrt{n}-z_{\xi_1})}{\sqrt{n-d}}}^{+\infty}\frac{1}{\sqrt{2\pi}\sigma}e^{-\frac{z_{\xi_2}^2}{2\sigma^2}}~{\rm
    d}z_{\xi_2},
\end{equation}
which is a strictly increasing and continuous function of $z_{\xi_1}$ such that $p_2(-\infty, d) = 0$ and $p_2(\sqrt{n}, d) = 1/2$. Then the conditional union bound is given by
\begin{equation}\label{TB_f_u}
  f_u(z_{\xi_1})=\sum_{d=1}^{n}A_{d} p_2(z_{\xi_1}, d),
 \end{equation}
which is also strictly increasing and continuous function of $z_{\xi_1}$ such that $f_u(-\infty) = 0$ and $f_u(\sqrt{n}) \geq 3/2$. Hence there exists a unique solution $z_{\xi_1}^* \leq \sqrt{n}$ satisfying

\begin{equation}\label{TB_opt}
  \sum_{d=1}^{n}A_{d} p_2(z_{\xi_1}, d) = 1,
 \end{equation}
which is equivalent to that given in~\cite[(3.22)]{Sason06} by noticing that $p_2(z_{\xi_1}, d) = Q\left(\frac{\sqrt{d}(\sqrt{n}-z_{\xi_1})}{\sigma\sqrt{n-d}}\right)$ and $d = \delta_d^2 / 4$.

\subsubsection{Equivalence}
The TB can be written as
\begin{eqnarray}\label{TB}
{\rm Pr} \{E\} &\leq& \int_{-\infty}^{z_{\xi_1}^*} \!\!\!
f_{u}(z_{\xi_1})  g(z_{\xi_1}) ~{\rm d}z_{\xi_1} + \int_{z_{\xi_1}^*}^{+\infty}
\!\!\!\!\!g(z_{\xi_1})~{\rm d}z_{\xi_1}\nonumber\\
&=&\int_{-\infty}^{+\infty} \min\{f_u(z_{\xi_1}), 1\} g(z_{\xi_1})~{\rm d}z_{\xi_1},
\end{eqnarray}
where $g_(z_{\xi_1})$ and $f_u(z_{\xi_1})$ are given in~(\ref{TB_g}) and~(\ref{TB_f_u}), respectively. The optimal parameter $z_{\xi_1}^*$ is given by solving the equation~(\ref{TB_opt}). It can be shown that~(\ref{TB}) is equivalent to that given in~\cite[(3.21)]{Sason06}.

\subsection{The Tangential-Sphere Bound Revisited}

Assume that $n\geq 3$.

\subsubsection{Nested Regions}
Again, the TSB chooses the nested regions to be a family of half-spaces $Z_{\xi_1} \leq z_{\xi_1}$, where $z_{\xi_1} \in \mathbb{R}$ is the parameter.

\subsubsection{Probability Density Function of the Parameter}

The pdf of the parameter is
\begin{equation}\label{TSB_g}
    g(z_{\xi_1}) = \frac{1}{\sqrt{2\pi}\sigma}e^{-\frac{z_{\xi_1}^2}{2\sigma^2}}.
\end{equation}

\subsubsection{Conditional Upper Bound}
Different from the TB, the TSB chooses $f_u(z_{\xi_1})$ to be the conditional sphere bound.
The conditional sphere bound given that  $Z_{\xi_1} = z_{\xi_1}$ can be derived as follows.

Let $\mathcal{R}(r)$ be the $(n-1)$-dimensional sphere of radius $r>0$ which is centered at $(1-z_{\xi_1}/\sqrt{n})s^{(0)}$ and located inside the hyper-plane $Z_{\xi_1} = z_{\xi_1}$.
\begin{itemize}
  \item[] {\em Case 1}: $Z_{\xi_1} = z_{\xi_1} \geq \sqrt{n}$.  It can be shown that, given that received vector falls on $\partial\mathcal{R}(r)$, the pair-wise error probability is no less than 1/2.   Hence the conditional union bound is no less than 3/2. From Theorem~\ref{Theorem_r1}, we know that the optimal radius $r_1(z_{\xi_1}) = 0$, which results in the trivial upper bound  $f_{u}(z_{\xi_1}) \equiv 1$.

  \item[] {\em Case 2}: Given that $Z_{\xi_1} = z_{\xi_1} < \sqrt{n}$, the ML decoding error probability can be evaluated by considering an equivalent system in which each bipolar codeword is scaled by a factor $(\sqrt{n} - z_{\xi_1})/\sqrt{n}$ before transmitted over an AWGN channel with~({\em projective}) noise $(0, Z_{\xi_2}, \cdots, Z_{\xi_n})$. The system is also equivalent to transmission of the original codewords over an AWGN but with scaled~({\em projective}) noise $\sqrt{n}/(\sqrt{n} - z_{\xi_1}) (0, Z_{\xi_2}, \cdots, Z_{\xi_n})$. The latter reformulation allows us to get the conditional sphere bound easily since the optimal radius is independent of the SNR. Actually, we notice that, given that the noise falls on the $(n-1)$-dimensional sphere $\partial \mathcal{R}(r)$ in the hyper-plane $z_{\xi_1} = 0$, the conditional pair-wise error probability is
        \begin{equation*}
            \!\!\!\!\!p_2(r, d)\!\! = \!\!\frac{\Gamma(\frac{n-1}{2})}{\sqrt{\pi}~\Gamma(\frac{n-2}{2})}\int_{0}^{\arccos(\frac{\sqrt{nd/(n-d)}}{r})}\sin^{n-3}
        \phi~{\rm d}\phi
        \end{equation*}
if $r > \sqrt{nd/(n-d)}$ and $p_2(r, d) = 0$ otherwise. Then we have the conditional sphere bound

       \begin{equation}\label{TSBu}
            f_u(z_{\xi_1}) = \int_{0}^{r_{1}}  f_{s}(r)g_{s}(z_{\xi_1}, r) ~{\rm d}r + \int_{r_{1}}^{+\infty}g_{s}(z_{\xi_1}, r)~{\rm d}r,
       \end{equation}
       where
       \begin{equation}\label{TSBu_g}
        g_s(z_{\xi_1}, r) = \frac{2r^{n-2}e^{-\frac{r^{2}}{2\tilde{\sigma}^{2}}}}{2^{\frac{n-1}{2}}\tilde{\sigma}^{n-1}\Gamma(\frac{n-1}{2})},~~r\geq 0,
        \end{equation}
        which depends on the SNR via $\tilde \sigma = \sqrt{n}\sigma/(\sqrt{n} - z_{\xi_1})$, and
        \begin{equation}\label{TSBu_f}
            \!\!\!\!f_s(r) = \!\!\!\!\!\sum_{1\leq d \leq \frac{r^2n}{r^2 + n}} A_d \frac{\Gamma(\frac{n-1}{2})}{\sqrt{\pi}~\Gamma(\frac{n-2}{2})}\int_{0}^{\arccos(\frac{\sqrt{nd/(n-d)}}{r})}\!\!\!\!\!\!\sin^{n-3}
            \phi~{\rm d}\phi,
        \end{equation}
    which is independent of $\tilde{\sigma}$, as justified previously. The optimal radius $r_1$ is the unique solution of
    \begin{equation}\label{TSBu_opt}
        \sum_{1\leq d \leq \frac{r^2n}{r^2 + n}} A_d \frac{\Gamma(\frac{n-1}{2})}{\sqrt{\pi}~\Gamma(\frac{n-2}{2})}\int_{0}^{\arccos(\frac{\sqrt{nd/(n-d)}}{r})}\!\!\!\!\!\!\sin^{n-3}
            \phi~{\rm d}\phi = 1.
    \end{equation}
Since  $r_1 < +\infty$, $f_u(z_{\xi_1}) < 1$ for all $z_{\xi_1} < \sqrt{n}$.

\item[]{\em Summary:} We have shown that the conditional sphere upper bound such that $f_u(z_{\xi_1}) < 1 $ if $z_{\xi_1} < \sqrt{n}$; otherwise, $f_u(z_{\xi_1}) = 1 $. Hence the optimal parameter $z_{\xi_1}^* = \sqrt{n}$.
\end{itemize}

\subsubsection{Equivalence} The TSB can be written as
\begin{eqnarray}\label{TSB}
{\rm Pr} \{E\}\!\!\! &\leq&\!\!\! \int_{-\infty}^{\sqrt{n}}\!\!\!\!\!
f_{u}(z_{\xi_1})  g(z_{\xi_1}) ~{\rm d}z_{\xi_1} + \int_{\sqrt{n}}^{+\infty}
\!\!\!\!\!g(z_{\xi_1})~{\rm d}z_{\xi_1},
\end{eqnarray}
where $g(z_{\xi_1})$ is given by~(\ref{TSB_g}), and $f_u(z_{\xi_1})$ is given  by~(\ref{TSBu})-(\ref{TSBu_opt}).

To prove the equivalence of~(\ref{TSB}) to the formulae given in
~\cite[Sec.3.2.1]{Sason06}, we first show that the optimal region is
the same\footnote{Strictly speaking, our derivations here show that the optimal region is a half-cone rather than a full-cone, a fact that has never been explicitly stated in the literatures. Once the optimal region is the same, the two
bounds should be the same except that they compute the bounds in
different ways.} as that given in~\cite[Sec.3.2.1]{Sason06}. Noting
that the optimal radius $r_1$ satisfies~(\ref{TSBu_opt}), which is
equivalent to that given in~\cite[(3.12)]{Sason06}. Back to the
hyper-plane $Z_{\xi_1} = z_{\xi_1}$, we can see that the optimal
parameter is $r_1 (\sqrt{n} - z_{\xi_1})/\sqrt{n}$. This means that
the optimal region is a half-cone with the same angle as that given
in~\cite[(3.12)]{Sason06}. Then, by changing variables, $r' = r
(\sqrt{n}-z_{\xi_1}) / \sqrt{n}$, $z_{\xi_2} = r'\cos\phi$, $v =
r'^{2}-z_{\xi_2}^{2}$ and $y = r'^{2}$, it can be verified
that~(\ref{TSB}) is equivalent to that given
in~\cite[(3.10)]{Sason06}, except that the second term ${\rm Pr}\{Z_{\xi_1} \geq \sqrt{n}\}$. This term did not appear in the original derivation of TSB in~\cite{TSB94}, but is required as pointed out in~\cite[Appendex A]{Sason00}.

\section{Conclusions}\label{conclusion}

In this paper, we have presented a general framework to investigate Gallager's first bounding technique with a single
parameter. We have presented a sufficient and necessary condition for the optimal parameter.
With the proposed general framework, we have re-derived three well-known bounds and presented the
relationships among them. We have also revealed a fact that the SB of
Herzberg and Poltyrev is equivalent to the SB of Kasami~{\em
et~al}.

%

\small
\bibliographystyle{IEEEtran}
\bibliography{IEEEabrv,tzzt}

\end{document}